\documentclass[11pt,a4paper]{article}
 
\usepackage[margin=1in]{geometry}
\usepackage{amsmath}
\usepackage{amssymb}
\usepackage{graphicx}
\usepackage{bm}
\usepackage{mathtools}
\usepackage{authblk}
\usepackage{cite}
\usepackage{hyperref}
\usepackage{multibib}
\newcites{supp}{References}
\usepackage{xr}
\usepackage{algorithm}
\usepackage{algpseudocode}

\title{\Large General \textit{in situ} feedback control of cascaded liquid crystal spatial light modulators for structured field generation}
\date{}
\author[1*$\dagger$]{An Aloysius Wang}
\author[1$\dagger$]{Yuxi Cai}
\author[1]{Zhenglin Li}
\author[1]{Ruofu Liu}
\author[1]{Yifei Ma}
\author[1]{Patrick S Salter}
\author[1*]{Chao He}
\affil[1]{Department of Engineering Science, University of Oxford, Parks Road, Oxford, OX1 3PJ, UK}
\affil[*]{Corresponding authors: aloysius.wang@gmail.com, chao.he@eng.ox.ac.uk}
\affil[$\dagger$]{These authors contributed equally to this work}

\begin{document}
\maketitle
{\bf Cascaded liquid crystal spatial light modulators provide a versatile strategy for the generation of structured light and matter fields, with applications including optical communications, photonic computing, and topological field engineering. However, experimental imperfections, such as temperature-dependent liquid crystal response, variations between individual pixels, and alignment errors, present significant engineering challenges in generating high-quality fields. Moreover, changes in experimental conditions over time mean that calibrating each component once is insufficient for maintaining long-term, high-quality field generation. To address this, we present a general engineering approach based on a bespoke, physically informed, and manifold-constrained gradient-descent scheme that enables \textit{in situ} feedback control, compensating for such errors in real time without the need to alter the experimental setup. We further demonstrate the correction efficacy of our proposed strategy through experiments in both spatially varying light and matter field generation, including scenarios in which complex vectorial aberrations are artificially introduced into the setup. Together, these demonstrations underscore the practicality of our method and its suitability for deployment in real-world experimental environments, paving the way for robust operation of cascaded architectures for structured field generation.}\\

The use of optical retarders \cite{BornWolf1999PrinciplesOfOptics}, such as liquid crystal devices \cite{Yang2023_LC_SLM_Review} and metasurfaces \cite{Ji2023MetasurfaceDesignQuantumOptics, balthasar_mueller_metasurface_2017, yu_broadband_2012}, as a means of phase control is a well-established strategy for structured beam generation \cite{He2022} and plays an integral role in many key applications, including holography and the generation of orbital angular momentum modes \cite{Genevet2015HolographicOpticalMetasurfaces, Forbes2016_CreationDetectionOpticalModes, Fang2020_OAM_Holography, Shen2019OpticalVortices}. However, the anisotropic nature of optical retarders means that they can also be used to control polarization state, and this capability has recently been developed into powerful techniques for generating complex structured polarization \cite{Moreno2012CompletePolarization, Han2013VectorialOptical, Rong2014GenerationArbitraryVector, Wang2007GenerationArbitraryVector, hu_arbitrary_2020, Converter} and matter fields \cite{he2023universal}, with the method already adopted in applications such as Stokes skyrmion generation \cite{wang2024topological, wang2024generalizedskyrmions}, vectorial adaptive optics \cite{Ma2025adv7904, Hu2021ArbitraryComplexRetarders, He2023VectorialAdaptiveOptics, Ma2024Vectorial, He2020_VectorialAdaptiveOptics}, polarization microscopy \cite{Deng2023_H_E_birefringence, Zhang2023_OptLett_6136, Shen2022_PolarizationAberrations}, photonic computing \cite{Wang2025PerturbationResilient}, and information storage \cite{chopsticks}.

In contrast to phase control, which can be achieved with a single spatially varying linear retarder with tunable retardance and a fixed axis, the greater dimensionality of polarization implies that, at minimum, a cascade of two such retarders is required for fixed-to-arbitrary polarization state generation, while a cascade of three is required to construct arbitrary elliptical retarder fields \cite{Zhang2025Elliptical} and thereby perform arbitrary-to-arbitrary polarization conversion \cite{he2023universal}. This use of cascades presents important engineering challenges not encountered in phase control. In particular, vectorial aberrations introduced by relay optics between layers and by waveplates used to adjust the polarization state, as well as alignment and angular errors, can compound in a cascaded configuration and significantly affect the quality of the generated fields \cite{He2023VectorialAdaptiveOptics}.

Moreover, different optical retarders also present with their own engineering challenges. Consider, for example, the commonly used liquid crystal spatial light modulators (LC-SLMs), which stand out for their commercial availability, pixel-wise control, reconfigurability, and dynamic range. Apart from ever-present systemic errors such as alignment and angular deviations, which inevitably affect the generated field, the temperature dependence of the device's retardance leads to time-varying changes in the output field as environmental conditions fluctuate, while pixel-to-pixel variations and hysteresis render pixel-wise calibration necessary yet challenging to perform \textit{in situ}. This is particularly so in SLM cascades, where component removal for calibration is both impractical and cumbersome. As such, the reliable generation of high-quality fields requires strategies capable of compensating for these variations in real time, a requirement that is crucial for ensuring consistent performance across a wide range of applications.

To address these issues, we propose a general \textit{in situ} feedback control strategy for tuning the phase patterns of a cascaded SLM system without requiring modifications to the experimental setup, thereby achieving high-quality structured field generation. We first address the problem of polarization field generation, before turning to the more complex problem of matter field generation, and present experimental results demonstrating the correction efficacy of our strategy in both cases. Lastly, we show that the feedback loop can compensate for artificially introduced complex aberrations added to the setup, thereby proving its practical applicability in real-world settings. To the best of our knowledge, this is the first demonstration of a strategy that simultaneously corrects phase patterns across an entire retarder cascade, enabling accurate and reliable structured field generation without requiring component removal for individual calibration.

\section*{Main}

The core idea underlying our strategy for both light and matter field generation is to approximate an otherwise inherently coupled optimization problem, involving millions of variables associated with the phase of each pixel in the SLM cascade, by a decoupled one that can be implemented through simple pixel-wise updates. The overall strategy is based on a bespoke, physically informed, manifold-constrained gradient-descent algorithm that iteratively corrects the phase levels of the individual SLMs to converge toward the target field. These modifications address two important difficulties that arise in the direct application of traditional gradient-descent methods \cite{Nocedal2006NumericalOptimization, Boyd2004ConvexOptimization}, which would otherwise render such approaches impractical. The first challenge is the sheer dimensionality of the problem, as a typical SLM contains on the order of a million pixels, each of which is an independent variable that must be optimized. The second challenge is that the target field, which is represented by vectors on the Poincar\'e sphere in the case of polarization field generation and by rotation matrices in the case of matter field generation, takes values in $S^2$ and $\mathrm{SO}(3)$, respectively. This introduces a nonlinear constraint \cite{Lee2013SmoothManifolds} that must be handled carefully to ensure the numerical stability of the algorithm.

We address these challenges one at a time, providing a qualitative overview in the main text and deferring technical details to the Methods section. Summaries of the overall strategy for polarization and matter field generation are also provided in Algorithms~\ref{alg:feedback} and~\ref{alg:matter}, respectively. Starting with the issue of dimensionality, as noted above, one effective approach is to seek an approximate, decoupled formulation of the problem, so that the overall optimization reduces to individual pixel-wise subproblems whose solutions can be obtained analytically. By doing so, a single large, coupled problem is replaced by a collection of independent, low-dimensional pixel-wise updates, making real-time feedback computationally feasible. Moreover, analytic per-pixel solutions eliminate the need for an iterative inner loop, which would otherwise substantially increase computational cost and hinder experimental implementation.

The primary complication in decoupling the problem arises from the cascade structure. In isolation, the appropriate use of relay optics implies that individual pixels of a fixed SLM affect, to good approximation, a disjoint set of camera pixels. We refer to this property as {\it approximate locality}. There is, however, no guarantee that these sets remain disjoint for different SLMs, which leads to coupling between pixels on different SLMs and, consequently, across the entire system. This motivates the use of a Gauss--Seidel method \cite{Bertsekas1999NonlinearProgramming} within each gradient descent iteration, in which the phase pattern on each SLM is updated sequentially while the patterns on the remaining SLMs are held fixed, thereby splitting the overall optimization into several steps that, individually, are well approximated by decoupled pixel-wise subproblems due to locality.

Next, to determine an appropriate step size in gradient descent without excessive evaluations of the loss function, each of which requires a single physical measurement \cite{He2022FullPoincarePolarimetry, Azzam16, Azzam78}, the pixel-wise subproblem can be further simplified by replacing the objective with a first-order local approximation and solving the resulting optimization exactly. At this stage, the nonlinear constraints become significant, as standard Taylor expansions do not generally respect this structure, leading to poor approximations and, consequently, unstable iterates unless the initial conditions are very close to optimal. To address this, we endow $S^2$ with its standard Riemannian structure and $\mathrm{SO}(3)$ with its standard Lie group structure, and perform Taylor expansions in the relevant tangent spaces using exponential charts, thereby producing nonlinear local approximations that obey the respective manifold-valued constraints \cite{Lee2013SmoothManifolds}.

An unfortunate consequence of using a nonlinear expansion is that the resulting pixel-wise subproblem also becomes nonlinear, and therefore difficult to solve in general. To address this, we once again exploit locality and replace key variables with their area-averaged values over the relevant camera pixels. The simple functional form of the exponential map then reduces the pixel-wise subproblem to an analytically solvable form. Taken together, the combination of Gauss--Seidel updates, first-order local expansion, and area averaging yields a practical algorithm that replaces the original complex, coupled optimization with efficient pixel-wise updates.

\begin{figure}[!t]
    \centering
    \includegraphics[width=\textwidth]{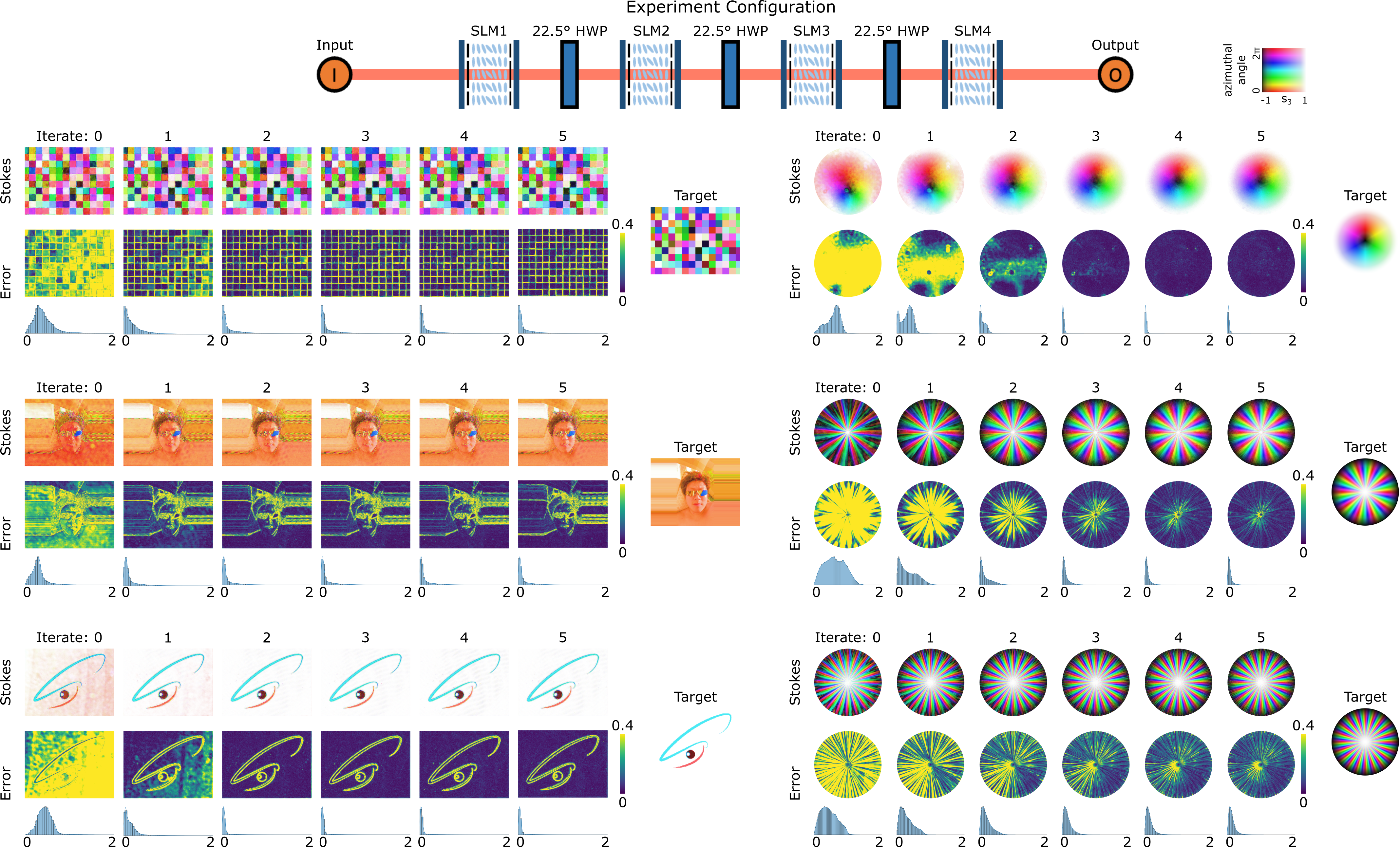}
    \caption{{\bf Experimental results (Light field generation).} Measured Stokes fields and corresponding $\ell^2$-error distributions (pixelwise and histogram) for various target Stokes fields across five feedback-loop iterations. For each histogram, the $x$-axis represents the $\ell^2$-error, while the $y$-axis is implicitly defined so that the total count across all bins equals the total number of camera pixels. A four-SLM cascade is used, with the relevant experimental configuration shown above. Throughout this paper, Stokes fields are depicted using color to represent the azimuthal angle on the Poincar\'{e} sphere and saturation to represent height \cite{shen_optical_2023}. (Top left) A lattice of randomly generated polarization states, which provides multiple analyzing channels and can thereby be used for applications such as one-shot polarization measurements. (Middle and bottom left) Color-encoded polarization images depicting a portrait and the Vectorial Optics and Photonics Group logo at the University of Oxford. (Right) Skyrmions of order 1, 10 and 20, respectively.}
\label{fig:experiment1}
\end{figure}

Having established the theoretical framework, we proceed to experimental demonstrations. Fig.\ \ref{fig:experiment1} shows the results of five feedback-loop iterations used to generate complex polarization fields for six independent examples, including a lattice of random polarization states for one-shot polarimetry, color-encoded polarization images exhibiting complex polarization structures, and Stokes skyrmions \cite{shen_optical_2023} of various orders. In our experimental setup, a cascade of four SLMs is used, with half-wave plates aligned at $22.5^\circ$ sandwiched between each SLM, while the phase decomposition is computed using the strategy presented in \cite{wang2024topological}. In this configuration, the dominant sources of error are expected to arise from angular and retardance errors in the SLMs and half-wave plates, which compound due to the cascaded architecture and result in an initially poor field.

The figure presents the experimentally measured Stokes field, together with the pixelwise $\ell^2$-error distribution and the corresponding histogram. For each histogram, the $y$-axis is implicitly normalized so that the total count across all bins equals the total number of camera pixels. From the figure, it is clear that in all experiments, the overall error is significantly reduced through feedback, and the strategy stabilizes at an optimum after only a few iterations of gradient descent (see Algorithm \ref{alg:feedback}). Note also that a single iteration requires only three Stokes measurements per SLM; thus, with a liquid-crystal polarimeter in which each Stokes measurement takes on the order of 1 ms, the measurement time for a single feedback step in a four-SLM cascade is approximately 10 ms, well within real-time operation. However, a few important caveats are worth mentioning. The first relates to a fundamental limitation of SLM cascades, namely that sharp changes in polarization state and high-frequency components of the target field cannot be realized, resulting in errors whenever such features are present. This behavior is clearly observed in Fig.\ \ref{fig:experiment1}, where a smooth, low-frequency field, such as an order-1 skyrmion, can be corrected to very high accuracy, while a field with many edges, such as the lattice, exhibits errors along these edges. The reasons for this are threefold, namely that the finite resolution of SLMs limits the spatial frequencies of the phase patterns they can produce, the continuous nature of the underlying liquid crystal structure prevents the implementation of discontinuous phase changes, and that diffraction naturally smooths out sharp changes in polarization state. Note that the latter also explains why the transfer function from SLM phase to polarization state is described as approximately local rather than strictly local. 

Next, the error distributions of the generated skyrmion fields shown in Fig.\ \ref{fig:experiment1} indicate that the norm condition for topological protection established in \cite{chopsticks} is satisfied with a large margin, ensuring that the topological number of the resulting field agrees with the target without the need to directly compute the skyrmion number integral. This demonstrates that feedback control enables the reliable generation of high-order Stokes skyrmions with the intended skyrmion number, opening practical routes to applications in optical communications and photonic computing \cite{Wang2025PerturbationResilient}.

\begin{figure}[!t]
    \centering
    \includegraphics[width=\textwidth]{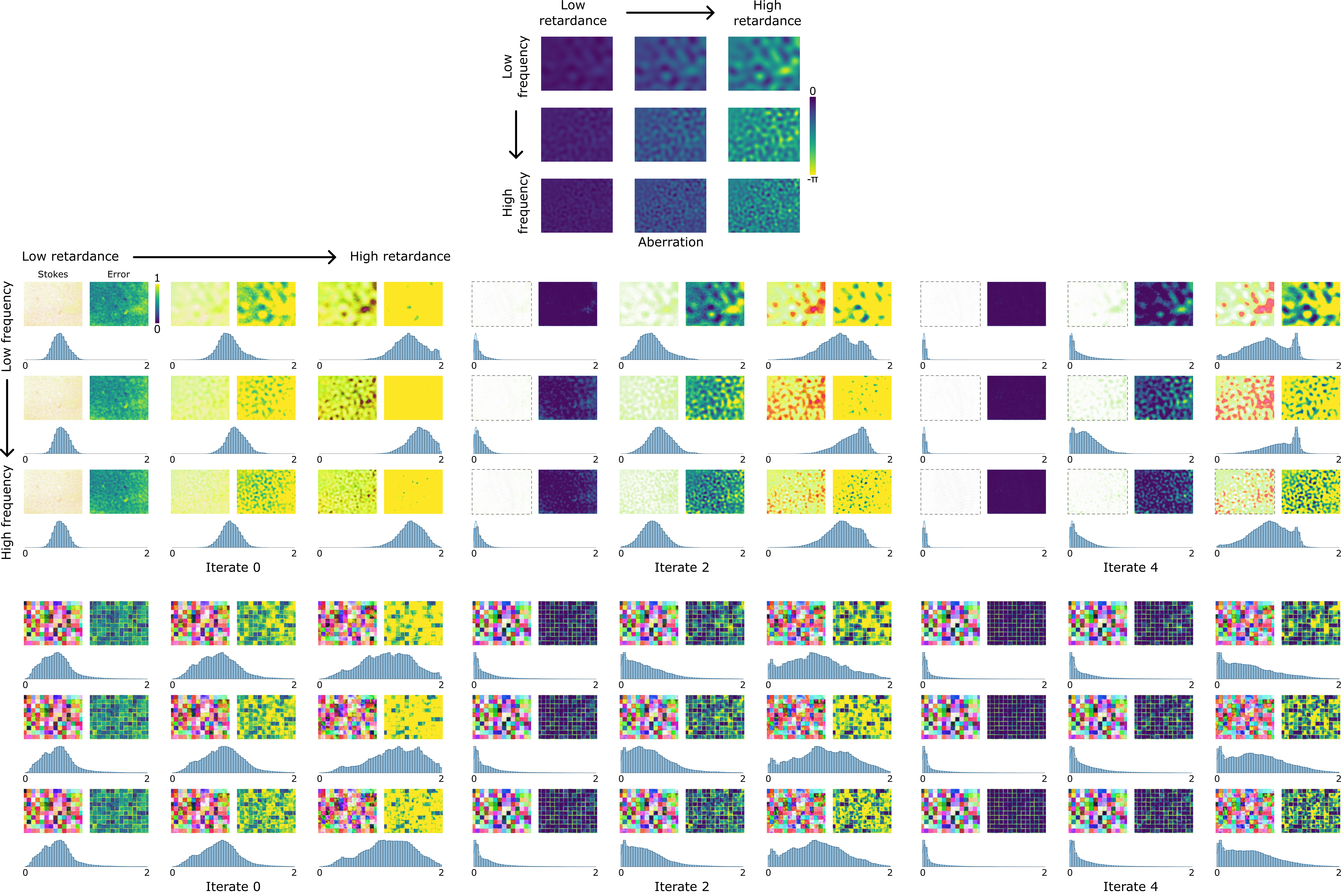}
    \caption{{\bf Experimental results (Light field generation with aberration).} Measured Stokes fields and corresponding $\ell^2$-error distribution (pixelwise and histogram) for various target Stokes fields across feedback-loop iterations under different polarization aberrations. For each target field, four feedback-loop iterations are performed (with iterations 0, 2, and 4 shown), while nine different polarization aberrations are considered, each characterized by a distinct maximum spatial frequency and maximum retardance. (Top) Phase pattern applied to the fourth SLM, mimicking a band-limited polarization aberration. (Middle) Experimental results for a target field that is uniformly right-circularly polarized. (Bottom) Experimental results for a target field consisting of a lattice of randomly generated polarization states, identical to that presented in Fig.\ \ref{fig:experiment1}}
\label{fig:experiment2}
\end{figure}

To further evaluate the effectiveness of the proposed feedback strategy, Fig.\ \ref{fig:experiment2} presents results obtained when a complex polarization aberration is introduced into the optical path. In this experiment, the cascade structure is unchanged, except that the final SLM is used to introduce a polarization aberration, while optimization is restricted to the first three SLMs. The aberrating phase patterns are generated as a band-limited superposition of plane waves with randomly generated phases and a spatial-frequency cutoff $k_{\text{max}}$,
\begin{equation*}
\phi(x,y) = a\Re\left[\sum_{\lvert \mathbf{k}\rvert< k_{\text{max}}}
e^{\mathrm{i}\left(k_x x + k_y y + \theta(\mathbf{k})\right)}\right]+b
\end{equation*}
where $\theta(\mathbf{k}) \sim \mathcal{U}(0,2\pi)$ is a randomly chosen phase offset and the constants $a$ and $b$ chosen so that $\phi(x,y)\in[-\sigma_{\text{max}}, 0]$. In our experiments, $\sigma_{\text{max}}$ takes values $\pi/6$, $\pi/2$ and $\pi$, representing external polarization aberrations of increasing strength that are relevant to practical optical systems. Two target fields are considered here over four iterations of feedback: a uniformly right-circularly polarized field and the same lattice of random polarization states as in Fig.\ \ref{fig:experiment1}.

From the figure, it is clear that for moderate polarization aberrations, the feedback strategy is able to correct for initial errors. However, as the magnitude of the aberration increases, the feedback process becomes progressively slower, which is expected behavior. Indeed, because the maximum phase correction applied to each SLM is deliberately limited to ensure the stability of the feedback algorithm (Methods 1.1), larger aberrations must be compensated through multiple incremental updates, resulting in slower overall convergence. If the initial aberration is too large, an additional factor comes into play that further slows the reduction in error: the local expansion used to compute pixel-wise updates is evaluated too far from the optimum for the resulting approximation to accurately reflect the true solution. As observed in the figure, although the error distribution improves in the presence of large polarization aberrations, the reduction in error per iteration is less pronounced than in the cases of small and moderate aberrations. In particular, when the retardance aberration is largest, the maximum error does not change significantly between the second and fourth iterations for a uniformly right-circularly polarized target field. Similarly, for the lattice of randomly generated polarization states, regions of maximum error persist even after four iterations of feedback. Together, these observations suggest that when the initial conditions are too far from optimality, the algorithm may become trapped in a local optimum, leading to stagnation of the optimization process---a behavior that is well known in gradient-type optimization strategies.

\begin{figure}[!t]
    \centering
    \includegraphics[width=\textwidth]{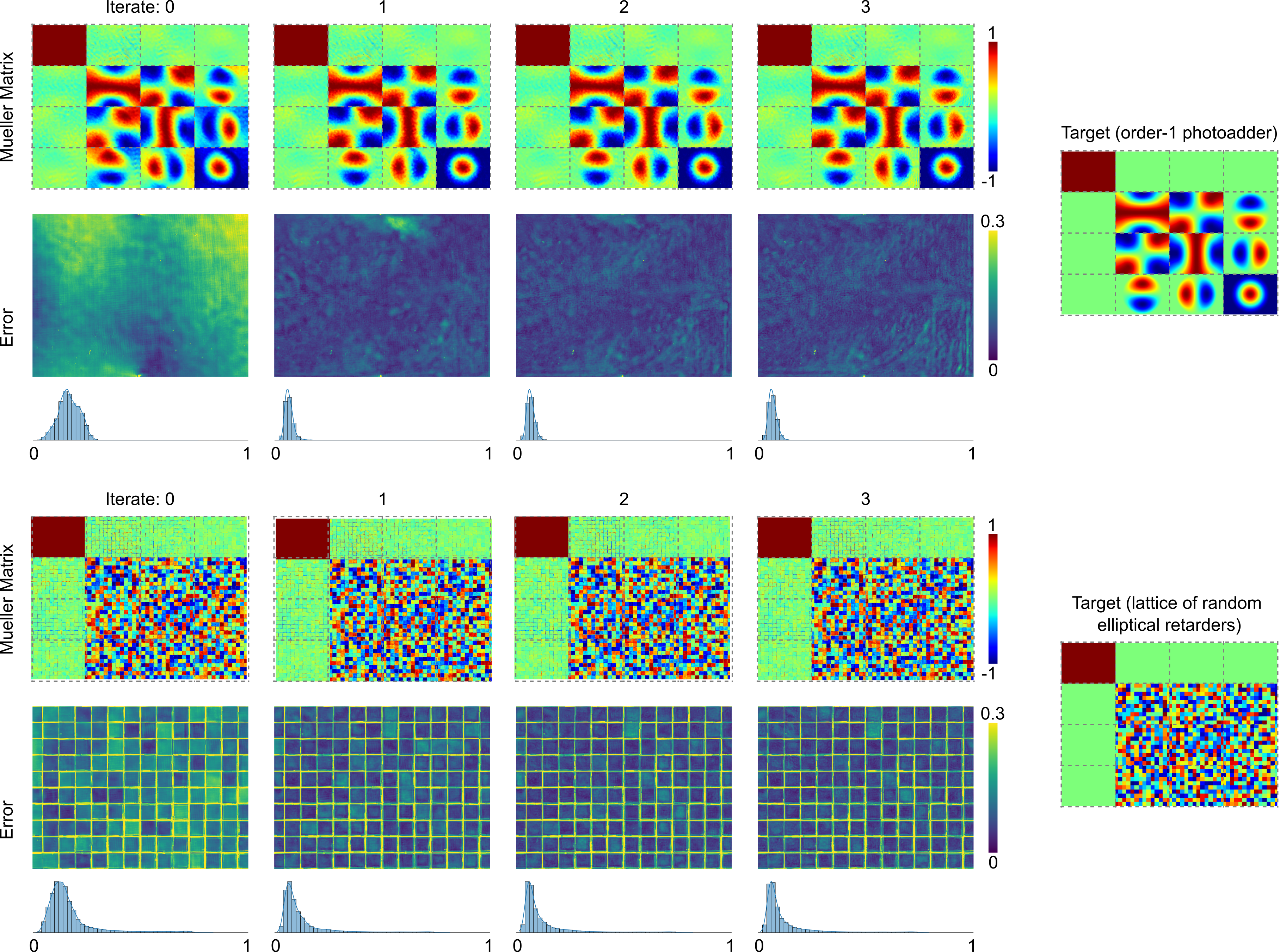}
    \caption{{\bf Experimental results (Matter field).} Measured Mueller matrices and corresponding Frobenius-norm error distributions (pixelwise and histogram) for various target matter fields across three feedback-loop iterations. (Top) An order-1 skyrmion photo-adder \cite{Wang2025PerturbationResilient} (Bottom) A lattice of random elliptical retarders.}
\label{fig:experiment3}
\end{figure}

Lastly, Fig.\ \ref{fig:experiment3} shows the results of three feedback-loop iterations used to generate complex matter fields. Two examples are given, namely an order-1 skyrmion photo-adder \cite{Wang2025PerturbationResilient} and a lattice of randomly generated elliptical retarders, using the same four-SLM cascade as in polarization field generation. In each case, the error distribution, measured by a normalized Frobenius norm $\tfrac{1}{4}\lVert \cdot \rVert_F$, is seen to decrease with feedback and quickly converge to a local optimum. Note, however, that at convergence the error distribution does not peak at zero, as is the case for Stokes measurements. This is because, in practice, the diattenuation vector (corresponding to $(m_{01}, m_{02}, m_{03})$) and polarizance vector (corresponding to $(m_{10}, m_{20}, m_{30})$) of the system are never perfectly zero \cite{Jorge2022}, and a small amount of depolarization is introduced by reflection from each SLM. Consequently, the measured Mueller matrix does not lie exactly on $\mathrm{SO}(3)$, and physical limitations of the system mean that this error cannot be eliminated. Nevertheless, as demonstrated by Fig.\ \ref{fig:experiment3}, the proposed feedback strategy produces a relatively good optimal solution and remains robust and convergent in practice.

Before concluding, we note two straightforward generalizations of our \textit{in situ} feedback correction approach. The first is that our proposed optimization strategy is \textit{independent} of cascade structure, choice of physical implementation, and the number of optical retarder layers used. Instead, it requires only reconfigurability of the devices and can therefore be extended to other optical-retarder-based systems, such as those employing metasurfaces \cite{Shaltout2019SpatiotemporalLight, Gu2023ReconfigurableMetasurfaces}. The second is that our proposed optimization strategy can easily be reformulated for target fields that lie in an arbitrary manifold $M$, and thus be extended to joint phase and polarization control, where $M = S^3$, as well as to full vectorial control, where $M = \mathbb{C}^2$ \cite{Zhao2025IntensityAdaptiveOptics}. This can be achieved, for example, in cascades of length greater than four and in systems incorporating a cross-polarizer for intensity control. Additionally, the strategy extends naturally to holographic settings \cite{Forbes2016_CreationDetectionOpticalModes, Fang2020_OAM_Holography}, which currently constitute the predominant approach to structured beam generation. Here, {\it approximate locality in $k$-space} is the key structural property enabling a similar approach. We leave the practical implementation of such systems as a possible extension of this work.

To conclude, in this work we present, for the first time, an {\it in situ} feedback correction strategy for light and matter field generation through a bespoke, physically informed, manifold-constraint gradient-descent scheme, providing a flexible and robust framework for high-fidelity structured-field generation in practical cascaded optical systems where componentwise calibration can be cumbersome or impractical. These results highlight the potential of the proposed approach to support a wide range of future structured-light applications, including advanced light and matter field control, adaptive optics, and emerging photonic technologies such as topological phases of light \cite{waveguide}. 

\clearpage

\bibliographystyle{naturemag}
\bibliography{main}

\clearpage

\section*{Methods}
\setcounter{section}{0}
\section{Feedback control strategy}
\subsection{Polarization field generation}
For polarization field generation, the calibration of an $n$-SLM cascade can be written as an optimization problem 
\begin{equation*}
    \min_{\phi_1,\ldots, \phi_n} \sum_{(x^\text{cam}, y^\text{cam})} \lVert f(\phi_1, \ldots, \phi_n)(x^\text{cam},y^\text{cam})-\mathcal{S}_\text{target}(x^\text{cam},y^\text{cam})\rVert^2
\end{equation*}
where $\phi_i \colon ([0, W_i-1]\times [0, H_i-1])\cap \mathbb{Z}^2 \longrightarrow [0, L_i-1] \cap \mathbb{Z}$ denotes the phase pattern applied to the $i$\textsuperscript{th} SLM, which has dimensions $W_i \times H_i$ and maximum signal level $L_i$, $(x^{\mathrm{cam}}, y^{\mathrm{cam}})$ are camera pixel coordinates, $\mathcal{S}_{\mathrm{target}}$ is the target polarization field, and $f$ denotes the unknown transfer function mapping the SLM phase patterns to the physically realized polarization field, which accounts for voltage- and temperature-dependent retardance, optical aberrations, angular errors, misalignments, and other experimental imperfections.

A straightforward but na\"{\i}ve approach to solving this problem is to employ gradient descent. More specifically, denoting the loss function by \(\mathcal{L}\), one can iteratively adjust the phase pattern according to
\begin{equation*}
    \phi_i[x^{\text{SLM}_i}, y^{\text{SLM}_i}] \gets 
    \phi_i[x^{\text{SLM}_i}, y^{\text{SLM}_i}] 
    - \alpha \frac{\partial \mathcal{L}}{\partial \phi_i[x^{\text{SLM}_i}, y^{\text{SLM}_i}]},
\end{equation*}
where $(x^{\text{SLM}_i}, y^{\text{SLM}_i})$ denote the SLM pixel coordinates, $\alpha > 0$ is the step size, and $\phi_i$ is implicitly clipped to lie in the range $[0, L_i-1]$ before being rounded to the nearest integer. However, the difficulties associated with this approach are twofold. The first is in approximating \begingroup \small
\begin{multline*}
    \frac{\partial \mathcal{L}}{\partial \phi_i[x^{\text{SLM}_i}, y^{\text{SLM}_i}]} = 2\sum_{(x^\text{cam}, y^\text{cam})}   \Bigl(f(\phi_1,\ldots, \phi_n)(x^\text{cam}, y^\text{cam})-\mathcal{S}_\text{target}(x^\text{cam}, y^\text{cam})\Bigr) \\ \cdot  \nabla_{\phi_i[x^{\text{SLM}_i}, y^{\text{SLM}_i}]} f(\phi_1,\ldots, \phi_n)(x^{\text{cam}}, y^{\text{cam}}),
\end{multline*} \endgroup
which requires at least one Stokes measurement for each individual pixel in the cascade to compute the gradient term. The second difficulty lies in determining an appropriate step size \(\alpha\). In standard approaches, \(\alpha\) is determined iteratively using a line-search algorithm, such as the backtracking Armijo line search \citesupp{Armijo1966_Minimization}; however, each iteration of the line search requires an evaluation of the loss function \(\mathcal{L}\), and hence an independent Stokes measurement. Together, these two factors contribute to a single iteration of gradient descent requiring millions of Stokes measurements, rendering the approach clearly impractical. The central idea of this paper is therefore to modify the standard gradient descent method such that the two issues identified above can be resolved using as few Stokes measurements as possible.

As mentioned in the main text, the key structural property of $f$ exploited to simplify the problem is that it is approximately local, in the sense that a spatially localized change in the phase pattern on a single SLM produces a change in the output polarization field that is approximately spatially localized. More specifically, there exist affine maps \(\mathcal{T}_i\colon \mathbb{R}^2 \longrightarrow \mathbb{R}^2\) for each SLM such that 
\begin{equation*}
    \operatorname{supp} (f(\phi_1, \ldots \phi_i ,\ldots, \phi_n) - f(\phi_1,\ldots, \psi_i, \ldots \phi_n)) \approx \mathcal{T}_i \operatorname{supp}(\phi_i-\psi_i),
\end{equation*}
and which can be easily determined experimentally by, for example, edge detection. Indeed, under perfect relay optics, the approximation above becomes exact, and the effect of the pixel \((x^{\text{SLM}_i}, y^{\text{SLM}_i})\) on \(f\) is confined to the set
\begin{equation*}
    \Omega_i[x^{\text{SLM}_i}, y^{\text{SLM}_i}]
    \coloneqq
    \mathcal{T}_i\!\left(
    [x^{\text{SLM}_i}-0.5, x^{\text{SLM}_i}+0.5]
    \times
    [y^{\text{SLM}_i}-0.5, y^{\text{SLM}_i}+0.5]
    \right).
\end{equation*}
Due to real-world experimental imperfections, this is no longer the case. Nonetheless, one expects the quantity
\begin{equation*}
    \sup_{(x^\text{cam}, y^\text{cam}) \notin \mathcal{T}_i \operatorname{supp} (\phi_i - \psi_i)} \lVert f(\phi_1, \ldots \phi_i ,\ldots, \phi_n)(x^\text{cam}, y^\text{cam}) - f(\phi_1,\ldots, \psi_i, \ldots \phi_n)(x^\text{cam}, y^\text{cam}) \rVert
\end{equation*}
to be small. Therefore, to good approximation, each SLM induces a partition of the camera pixels into disjoint sets, such that within each set the output $f$ depends primarily on a single pixel of that SLM. By exploiting this, the resulting gradients become spatially sparse, as
\begin{equation*}
    \nabla_{\phi_i[x^{\text{SLM}_i}, y^{\text{SLM}_i}]} f(\phi_1,\ldots, \phi_i, \ldots, \phi_n)(x^{\text{cam}}, y^{\text{cam}}) \approx 0\text{, for }(x^\text{cam}, y^{\text{cam}})\notin \Omega_i[x^{\text{SLM}_i}, y^{\text{SLM}_i}],
\end{equation*}
and are approximately decoupled in space. This allows us to approximate, using only two measurements, \(\nabla_{\phi_i[x^{\mathrm{SLM}_i}, y^{\mathrm{SLM}_i}]} f(\phi_1, \ldots, \phi_i, \ldots, \phi_n)(x^{\mathrm{cam}}, y^{\mathrm{cam}})\) simultaneously for every pixel of the $i$\textsuperscript{th} SLM. In particular, this enables the efficient approximation of the partial derivatives of the loss function $\mathcal{L}$, thereby resolving the first major problem.

With access to first-order information, a common approach to addressing the second issue raised above is to solve the linearized problem exactly at each iteration, rather than evaluating the loss function iteratively within a line search. Specifically, at iteration $m$, the loss function can be approximated via \begingroup \small
\begin{multline*}
    f(\phi_1,\ldots,\phi_n)(x^\text{cam}, y^\text{cam}) \approx f(\phi_1^{(m)}, \ldots, \phi_n^{(m)})(x^\text{cam}, y^\text{cam})+\\ \sum_{i=1}^n \sum_{(x^{\text{SLM}_i}, y^{\text{SLM}_i})} (\phi_i[x^{\text{SLM}_i},y^{\text{SLM}_i}]-\phi_i^{(m)}[x^{\text{SLM}_i},y^{\text{SLM}_i}])\nabla_{\phi_i[x^{\mathrm{SLM}_i}, y^{\mathrm{SLM}_i}]} f(\phi_1^{(m)}, \ldots, \phi_n^{(m)})(x^{\mathrm{cam}}, y^{\mathrm{cam}}),
\end{multline*}
\endgroup
and the corresponding optimization problem is a linear least-squares problem, for which standard methods can be applied. Note that this approach is effective only when the region over which the local linear model is accurate contains the true optimum, and a poor initial condition that is far from optimality can lead to unstable updates. Despite this, the method remains feasible but leaves room for improvement in two key aspects. Firstly, while $f$ itself takes values in $S^2$, this structure is not preserved under linearization, which leads to the approximation being poor except at points very close to $(\phi_1^{(m)}, \ldots, \phi_n^{(m)})$. Secondly, the resulting optimization problem is extremely high-dimensional, and therefore computationally intractable to solve directly.

To address the first point, we endow $S^2$ with its standard Riemannian structure. The corresponding exponential map $\exp_p \colon T_p S^2 \to S^2$ is then a local diffeomorphism in a neighborhood of $0 \in T_p S^2$, allowing us to perform Taylor expansion in $T_p S^2$ instead. This yields a local expansion which respects the $S^2$ structure given by \begingroup \small
\begin{equation*}
    f(\phi_1,\ldots,\phi_n)(x^\text{cam}, y^\text{cam}) \approx \exp_{f(\phi_1^{(m)}, \ldots, \phi_n^{(m)})(x^\text{cam}, y^\text{cam})}(df_{(\phi_1^{(m)}, \ldots, \phi_n^{(m)})}(\phi_1-\phi_1^{(m)}, \ldots, \phi_n-\phi_n^{(m)})(x^\text{cam}, y^\text{cam})),
\end{equation*} \endgroup
where $df$ can be estimated experimentally by projecting $\nabla_{\phi_i[x^{\text{SLM}_i}, y^{\text{SLM}_i}]} f(\phi_1^{(m)},\ldots,\phi_n^{(m)})(x^{\text{cam}},y^{\text{cam}})$ onto the relevant tangent space. 

To address the second point, our strategy is to exploit the approximate locality of $f$, allowing the problem to be decoupled into a pixel-wise subproblem that can be solved analytically. The main challenge is that, although the sets $\Omega_i$ form a partition of the camera pixels for fixed $i$, there is no guarantee $\Omega_i[x^{\mathrm{SLM}_i}, y^{\mathrm{SLM}_i}]$ and $\Omega_j[x^{\mathrm{SLM}_j}, y^{\mathrm{SLM}_j}]$ are mutually disjoint. As mentioned in the main text, this can be resolved by applying the Gauss--Seidel method. In this case, the optimization corresponding to the $i$\textsuperscript{th} SLM at iteration $m$ reduces to
\begin{equation*}
    \phi_i^{(m+1)} = \arg\min_{\phi_i} \sum_{(x^\text{cam}, y^\text{cam})} \lVert f(\phi_1^{(m+1)},\ldots, \phi_i, \ldots, \phi_n^{(m)})(x^\text{cam},y^\text{cam}) -\mathcal{S}_\text{target}(x^\text{cam},y^\text{cam})\rVert^2,
\end{equation*}
so that, rewriting the sum in the loss function as 
\begin{equation*}
    \sum_{(x^\text{cam}, y^\text{cam})} = \sum_{(x^{\text{SLM}_i}\, y^{\text{SLM}_i})} \sum_{(x^\text{cam}, y^\text{cam}) \in \Omega_i[x^{\text{SLM}_i}\, y^{\text{SLM}_i}]}
\end{equation*}
and applying approximate locality, one obtains a fully decoupled approximate problem \begingroup \small
\begin{equation*}
    \phi_i^{(m+1)}[x^{\text{SLM}_i},y^{\text{SLM}_i}] = \arg\min_\phi \sum_{(x^{\text{cam}}, y^{\text{cam}}) \in \Omega_i[x^{\text{SLM}_i},y^{\text{SLM}_i}]} \lVert \exp_{p(x^\text{cam}, y^{\text{cam}})}((\phi-\phi_i^{(m)}) v(x^\text{cam}, y^{\text{cam}}))-\mathcal{S}_\text{target}(x^\text{cam},y^\text{cam})\rVert^2
\end{equation*} \endgroup
where \begingroup \small
\begin{align*}
    p&= f(\phi_1^{(m+1)}, \ldots, \phi_i^{(m)},\ldots, \phi_n^{(m)}),\\
    v &= \nabla_{\phi_i[x^{\text{SLM}_i}, y^{\text{SLM}_i}]} f(\phi_1^{(m+1)}, \ldots, \phi_i^{(m)},\ldots, \phi_n^{(m)}) - \left(p \cdot \nabla_{\phi_i[x^{\text{SLM}_i}, y^{\text{SLM}_i}]} f(\phi_1^{(m+1)}, \ldots, \phi_i^{(m)},\ldots, \phi_n^{(m)})\right)p.
\end{align*} \endgroup
Note that for clarity, the dependence on $(x^{\text{cam}}, y^{\text{cam}})$, as well as other variables, is omitted. One final simplification is to replace the quantities $p$, $v$, and $\mathcal{S}_{\text{target}}$ by their averages over $\Omega_i[x^{\text{SLM}_i}, y^{\text{SLM}_i}]$, which we denote by an overbar. This yields
\begin{equation*}
    \phi_i^{(m+1)}[x^{\text{SLM}_i}, y^{\text{SLM}_i}]
    =
    \arg\min_{\phi}
    \left\lVert
    \exp_{\bar{p}}((\phi-\phi_i^{(m)}) \bar{v})
    -
    \bar{\mathcal{S}}_{\text{target}}
    \right\rVert^2,
\end{equation*}
which can be solved explicitly as 
\begin{equation*}
    \phi_i^{(m+1)}[x^{\text{SLM}_i}, y^{\text{SLM}_i}]
    = \phi_i^{(m)}+\frac{1}{\lVert \bar{v}\rVert} \operatorname{atan2}\left(\frac{\bar{v}}{\lVert \bar{v}\rVert} \cdot \bar{\mathcal{S}}_{\text{target}}, \bar{p} \cdot \bar{\mathcal{S}}_{\text{target}}  \right). 
\end{equation*}

Letting $\langle \cdot \rangle_\Omega$ denote averaging over the camera pixels in $\Omega$, the overall feedback strategy is summarized in Algorithm \ref{alg:feedback}. Note that further practical modifications can be made to improve the reliability of the feedback strategy, including limiting each phase update to a maximum step length and smoothing the phase distribution to prevent high-frequency artefacts from appearing.

\begin{algorithm}
\caption{Polarization Field Feedback Strategy}
\label{alg:feedback}
\begin{algorithmic}[1]
\State Determine $\mathcal{T}_1,\ldots, \mathcal{T}_n$ by edge detection
\State Compute the sets $\Omega_i[x^{\text{SLM}_i}, y^{\text{SLM}_i}]$
\State Initialize the phase patterns $(\phi_1, \ldots, \phi_n)$
\For{$m = 1$ to $M$} \Comment{Gradient descent}
    \For {$i = 1$ to $n$} \Comment{Gauss--Seidel}
        \State Measure $\mathcal{S} \gets f(\phi_1,\ldots,\phi_i,\ldots,\phi_n)$
        \State Measure $\mathcal{S}_+ \gets f(\phi_1,\ldots,\phi_i+\delta\phi,\ldots,\phi_n)$
        \State Measure $\mathcal{S}_- \gets f(\phi_1,\ldots,\phi_i-\delta\phi,\ldots,\phi_n)$
        \For {$(x, y) \in ([0, W_i-1]\times [0, H_i-1])\cap \mathbb{Z}^2$} \Comment{Loop over SLM pixels}
            \State $\bar{p} \gets \langle \mathcal{S} \rangle_{\Omega_i[x,y]}$
            \State $\bar{p} \gets \bar{p}/\lVert \bar{p}\rVert$
            \State $\bar{g} \gets \frac{1}{2\delta\phi} \langle \mathcal{S}_+-\mathcal{S}_-\rangle_{\Omega_i[x,y]}$
            \State $\bar{v} \gets \bar{g} - (\bar{p} \cdot \bar{g})\bar{p}$
            \State $\phi_i[x,y] \gets \phi_i[x,y]+\operatorname{atan2}(\cdots)/{\lVert \bar{v}\rVert}$ \Comment{Phase update}
            \State $\phi_i[x, y] \gets \operatorname{round}(\operatorname{clip}(\phi_i[x, y], 0, L_i-1))$
        \EndFor
    \EndFor
\EndFor
\end{algorithmic}
\end{algorithm}

\subsection{Matter field generation}

In matter field generation, the above strategy can be adapted directly with only two small modifications. The first follows from the fact that, in this case, the unknown transfer function $f$ maps only approximately into $\mathrm{SO}(3)$ due to diattenuation, polarizance and depolarization, as explained in the main text. Thus, the nonlinear constraint obeyed by $f$ is itself unknown, and must be approximated by projecting $f$ onto $\mathrm{SO}(3)$ in a suitable manner, for example via least squares. The second difference arises from the functional form of the exponential map. In the case where the standard exponential map on $\mathrm{SO}(3)$ is used, and retaining the variable names from Methods~1.1 with their obvious definitions, one can write
\begin{equation*}
    \Omega = (L_{\bar{p}^T})_\ast \bar{v} \in \mathfrak{so}(3),
\end{equation*}
so that 
\begin{equation*}
    \exp_{\bar{p}}((\phi-\phi_i^{(m)})\bar{v}) = \bar{p} \left(I + \frac{\sin ((\phi-\phi_i^{(m)})\lVert \Omega \rVert)}{\lVert \Omega \rVert} \Omega + \frac{1-\cos((\phi-\phi_i^{(m)})\lVert \Omega \rVert)}{\lVert \Omega \rVert^2}\Omega^2\right),
\end{equation*}
where $\lVert \cdot \rVert$ is the canonical bi-invariant metric on $\mathrm{SO}(3)$. From this, the optimization problem
\begin{equation*}
    \arg\min_{\phi}
    \left\lVert
    \exp_{\bar{p}}((\phi-\phi_i^{(m)}) \bar{v})
    -
    \bar{\mathcal{M}}_{\text{target}}
    \right\rVert^2
\end{equation*}
can be solved directly, albeit with considerable effort. 

Note, however, that a slightly more straightforward strategy for matter field feedback, which does not require the above optimization to be solved at all, is to adopt the feedback correction for Stokes field generation repeatedly for various polarization state generator (PSG) configurations. More specifically, for $K$ different configurations that produce incident fields $\mathcal{S}_{\text{in}}^1, \ldots, \mathcal{S}_{\text{in}}^K$, a single iteration of the Stokes feedback algorithm can be applied, where at the $k$th step $\mathcal{S}_{\text{target}}^k = \mathcal{M}_{\text{target}} \mathcal{S}_{\text{in}}^k$. This strategy is more efficient than directly using the Mueller matrix as an input to feedback control, as it updates the phases after every four measurements rather than sixteen and implicitly accounts for nonlinear constraints; however, the optimal choice of $\mathcal{S}_{\text{in}}^k$ can be difficult to determine \emph{a priori}. A good rule of thumb is to ensure that the generated states $\mathcal{S}_{\text{in}}^k$ probe each entry of $\mathcal{M}_{\text{target}}$ with equal weight, which amounts to having a well-conditioned PSG instrument matrix \citesupp{Yuxi}. We leave a more detailed analysis for future work. The overall matter field feedback algorithm can thus be summarized in Algorithm \ref{alg:matter}.

\begin{algorithm}
\caption{Matter Field Feedback Strategy}
\label{alg:matter}
\begin{algorithmic}[1]
\For {$l=1$ to $L$}
    \For {$k=1$ to $K$}
        \State Set PSG to configuration $k$
        \State Apply a single loop of the Stokes field feedback strategy with $\mathcal{S}_\text{target} = \mathcal{M}_{\text{target}} \mathcal{S}_{\text{in}}^k$.
    \EndFor
\EndFor
\end{algorithmic}
\end{algorithm}

\bibliographystylesupp{naturemag}
{\bibliographysupp{main}}
\end{document}